\documentclass[prl,twocolumn,showpacs,amsmath,amssymb,superscriptaddress]{revtex4}

\usepackage{graphicx}% Include figure files
\usepackage{amssymb,amsmath}
\usepackage{dcolumn}% Align table columns on decimal point
\usepackage{bm}% bold math
\usepackage{color}

\begin{document}

\title{Skyrmion Formation and Optical Spin-Hall Effect\\ in an Expanding Coherent Cloud of Indirect Excitons}
\author{D. V. Vishnevsky}
\affiliation{Institut Pascal, PHOTON-N2, Clermont Universit\'{e}, Blaise Pascal University, CNRS, 24 avenue des Landais, 63177 Aubi\`{e}re Cedex, France.}
\author{H. Flayac}
\affiliation{Institut Pascal, PHOTON-N2, Clermont Universit\'{e}, Blaise Pascal University, CNRS, 24 avenue des Landais, 63177 Aubi\`{e}re Cedex, France.}
\author{A. V. Nalitov}
\affiliation{Institut Pascal, PHOTON-N2, Clermont Universit\'{e}, Blaise Pascal University, CNRS, 24 avenue des Landais, 63177 Aubi\`{e}re Cedex, France.}
\author{D. D. Solnyshkov}
\affiliation{Institut Pascal, PHOTON-N2, Clermont Universit\'{e}, Blaise Pascal University, CNRS, 24 avenue des Landais, 63177 Aubi\`{e}re Cedex, France.}
\author{N. A. Gippius}
\affiliation{Institut Pascal, PHOTON-N2, Clermont Universit\'{e}, Blaise Pascal University, CNRS, 24 avenue des Landais, 63177 Aubi\`{e}re Cedex, France.}
\affiliation{A. M. Prokhorov General Physics Institute, RAS, Vavilova Street 38, Moscow 119991, Russia.}
\author{G. Malpuech}
\affiliation{Institut Pascal, PHOTON-N2, Clermont Universit\'{e}, Blaise Pascal University, CNRS, 24 avenue des Landais, 63177 Aubi\`{e}re Cedex, France.}

\begin{abstract}
We provide a theoretical description of the polarization pattern and phase singularities experimentally evidenced recently in a condensate of indirect excitons [H. High et al., Nature \textbf{483}, 584-588 (2012)].
We show that the averaging of the electron and hole orbital motion leads to a comparable spin-orbit interaction for both type of carriers. We demonstrate that
the interplay between a radial coherent flux of bright indirect excitons and the Dresselhaus spin-orbit interaction results in the formation of spin domains and of topological defects similar to Skyrmions. We reproduce qualitatively all the features of the experimental data and obtain polarization pattern as in the optical spin Hall effect despite the different symmetry of the spin-orbit interactions.
\end{abstract}

\maketitle

\emph{Introduction}.
The recent implementation of synthetic gauge fields acting on spinor atomic Bose Einstein condensates \cite{Nature_Gauge_field} has opened a new direction of research \cite{review_dalibard}.
Indeed, the coupling between the spin degree of freedom and the motion of bosons, linear in $k$, completely modifies the nature of the condensed states. The bosons do not
accumulate any more in the single $k=0$ state, and the condensate can demonstrate spin-dependent modulations, such as stripes \cite{Sinha}. In confined systems, skyrmions and half vortices are expected to appear, and not as the excitations of the condensate, but directly in the ground state \cite{StableSkyrmions}. The exact nature of these original equilibrium phases depends both on the type of spin-orbit coupling and on the inter-particle interactions.

Indirect excitons in coupled quantum wells are bosonic particles which continue to attract a strong interest since the Lozovik and Yudson's proposal to use a condensate of such particles towards a new type of superconductivity \cite{Lozovik1}. The spin structure of indirect excitons is especially interesting and unique since, differently from bosonic atoms, they possess an even number of allowed spin projections on the sample's growth axis. Two spin projections $\pm1$ can couple to light to form bright states, while the $\pm2$ components are bound to be dark. Furthermore, both the electrons and the holes are affected by the Rashba and Dresselhaus spin-orbit interaction (SOI) which leads to a variety of accessible phases, both in the homogeneous \cite{ShelykhIndEx2012} and in the confined systems \cite{RuboKavokin2012}.

However, the experimental phenomenology of indirect exciton condensation is complex and still not completely understood. Their principal advantage with respect to direct excitons or exciton-polaritons is the reduced overlapping of their wave-functions that infers the quasi-particles a long lifetime in the range of nanoseconds. This feature is expected to optimize the process of thermalization of excitons which in turn should make the BEC formation easier. On the other hand, the critical condensation temperature is of about one Kelvin, a temperature at which an electron or an exciton gas is completely localized by the structural disorder present in any semiconductor structure. Moreover, only a very tiny fraction of the exciton distribution can be probed by optical means. The dark states cannot be detected, and the bright ones are allowed to emit only within the light cone which represents a very narrow part of the reciprocal space. Despite these apparent difficulties, some interesting experimental results have been recently obtained in Ref.\cite{ButovNat2012} and can be summarized as follows. Far away from the local pumping area, strong maxima in the bright exciton distribution are observed. They were attributed to complex processes of carrier diffusion. These localised bright spots (LBS) were found to generate a radial flow of bright excitons characterized by: 1) a spatial coherence, with a coherence length of the order of 10 microns; 2) a specific linear polarization pattern over the same length scale; and 3) the presence of phase singularities.
This specific linear polarization pattern has been attributed to the radial cloud expansion from localized sources (See supplemental material of Ref.\cite{ButovNat2012}). The wave vector distribution of the electron and the hole wavefunctions forming the excitonic wave packet was neglected. This approach allowed to reproduce the polarization pattern, but not the presence of phase singularities observed.

In this Letter, we consider a different and more complete approach, allowing to qualitatively reproduce all the experimental features. First, we take into account the internal structure of excitons and show how the relative motion of electron and hole affects the SOI. Especially, the hole SOI is found to become comparable to that of the electron. Second, we assume that a bright indirect exciton condensate with a well defined spin state is locally formed at the center of the LBS, as assumed in Ref.\cite {ButovNat2012}. Then, unlike the approach of Ref.\cite {ButovNat2012}, we consider the coherent expansion of the resonantly created exciton cloud by direct numerical solution of the Schroedinger equation, in the presence of the Dresselhaus SOI. The renormalization of the dispersion induces a radial flow of excitons outwards from the pump spot. The repulsive exciton-exciton interactions can also contribute to this effect, but we neglect them in the present work. Since the typical time scale of the scattering of indirect excitons on phonons  is of about a few ns \cite{PhysRevLett.86.5608}, the ballistic propagation length is expected to be of the order of 10-20 microns. Within this length scale, we reproduce both the polarization pattern, and the presence of phase singularities of the wave function components, which are associated with the formation of Skyrmions. These topological defects appear thanks to the interplay between the radial flow and the SOI, as recently shown theoretically \cite{NOSHE} for cavity polaritons flowing in a TE-TM effective magnetic field. Inspired by the experimental results, we propose a configuration leading to the onset of circular polarization domains, fully equivalent to the one observed in the optical spin-Hall effect\cite{OSHEth}. This happens despite the fact that the wavevector dependence of the effective magnetic field of the Dresselhaus SOI is completely different from the one given by the TE-TM splitting in the microcavities.

\emph{The model}.
Indirect excitons are composite bosons formed from electrons and heavy holes separated in space. Condensates of indirect excitons were observed in systems of double quantum wells \cite{Filin,BilayerBEC}, where the electrons and the holes were localized in different layers, and in wide single quantum wells under applied or internal electric fields \cite{Gorbunov}. The projection of the electron's spin (angular momentum of heavy hole) on the growth axis can take two values $\pm 1/2$ $(\pm 3/2)$, so the total angular momentum of exciton can take four values $\pm 2$ and $\pm1$. The bright $\left|\pm1\right\rangle$  states are coupled with $\sigma^{\pm}$-polarized light, while the states $\left|\pm2\right\rangle$ are dark. Their radiative recombination is forbidden by selection rules.

In planar structures these four states can be coupled via the inter-exciton interactions, the application of an external magnetic field, or by spin-orbit interactions (SOI) of carriers originating from the violation of the inversion symmetry. The interactions of the first type are density-dependent. There are two kinds of spin-orbit coupling for carriers: Dresselhaus \cite{Dresselhaus1955} and Rashba \cite{Rashba1984} SOIs. Both interactions act as effective in-plane $k$-dependent magnetic fields coupling different carrier spin states \cite{bookIvchenko}. However, in the sample used in the experiments of reference \cite{ButovNat2012, RuboKavokin2012, LeonardSpinTransport}, the Dresselhaus SOI has been found to be much larger than the Rashba SOI, and thus the latter shall be neglected in the following. %If it is not neglected, its action would actually be to slightly rotate the effective in-plane magnetic field and therefore the polarization domains.

For a gas of free 2D electrons, Dresselhaus SOI is linear versus the electron wave-vector and the corresponding term of the Hamiltonian in the basis $(+1/2,-1/2)^T$ takes the form:
\begin{equation}
H_{e}^{D}  = \tilde\beta _{e} (\sigma_+ k^e_- + \sigma_- k^e_+ ),
\label{DresHamEl}
\end{equation}
where $\tilde\beta_{e}$ is a Dresselhaus interaction constant. Here and after we use the following notations: $k^{\gamma}_{\pm} = k^{\gamma}_x \pm i k^{\gamma}_y$, $k^{\gamma}_{x,y}$ are $x$- and $y$-components of wave-vector $k^{\gamma}$, and $\gamma=$e, h or X for electrons, holes or excitons respectively. $\sigma_\pm = \sigma_x \pm i \sigma_y$ when $\sigma_{x,y}$ are Pauli matrices.

For free heavy holes in 2D, SOI terms are cubic versus the wave-vector $k$, and in the basis $(+3/2,-3/2)^T$ have the following form \cite{BulaevLoss}:
\begin{subequations}
\begin{equation}
{H_h^{\rm {R}}} = i \tilde \alpha_{h} ({\sigma _ + } k_-^3   - {\sigma _ - }k_+^3)
\label{FreeHole}
\end{equation}
\begin{equation}
{H_h^{\rm {D}}} = - \tilde \beta_{h} ({\sigma _ + }{k_- }{k_+}{k_-} + {\sigma _ - }{k_+ }{k_-}{k_+})
\end{equation}
\label{FreeHole}
\end{subequations}
Here $\tilde \alpha_{h}$ and $\tilde \beta_{h}$ are interaction constants. Due to the cubic dependence on $k$, SOI of holes was considered for a long time as a negligible mechanism of hole spin relaxation \cite{DyakonovSpinPhysics}. In latest studies \cite{ZungerPRL2010}, a splitting linear in $k$ between heavy hole states has been predicted considering the interactions of heavy-holes with remote energy states in the valence band. However, to our knowledge, nobody has yet considered how the Hamiltonian (\ref{FreeHole}) is transformed for a hole bound inside an exciton. To do this, we decouple the relative motion of electron and holes from  the motion of the excitonic center of mass:
\begin{equation}
\Psi(k^X,q) = \Psi_{\rm{c.m.}} (k^X) \Psi_{\rm{rel}} (q).
\end{equation}
Here $q$ is large and rapidly varying wave-vector, describing e-h relative motion and  $k^{X}$ relies to hole wave-vector as $k^{h}=\nu_hk^{X}$ where $\nu_{e,h}={{m_{e,h} }}/({m_e  + m_h })$ is the ratio of the electron (hole) mass to the exciton mass. Then averaging Hamiltonian (\ref{FreeHole}) over the relative motion ground state gives rise to Dresselhauss SOI linear in $k^h$ term for holes:
\begin{equation}
{H_h^D} = \int \vert \Psi_{\rm{rel}}(q) \vert^2 H_h^D (k^h + q) d^2 q = \beta_h (\sigma_+ k^h_- + \sigma_- k^h_+ ),
\label{BoundHole}
\end{equation}
while the Rashba term under similar averaging remains cubic in $k^h$ and can be anyway safely neglected. Here $\beta_h = - 2 \tilde\beta_h \left\langle q^2 \right\rangle $ is the effective Dresselhaus constant for holes. Note that $\left\langle q^2 \right\rangle \sim a_{\rm B}^{-2} \gg (k^h)^2$, where $a_{\rm B}$ is the excitonic Bohr radius. Analogous procedure for electrons does not change the form of (\ref{DresHamEl}) but renormalizes electron interaction constants. Finally, one may write a SOI term for both electron and hole bound into exciton as:
\begin{equation}
H_{e,h}^{D}  = \beta_{e,h} \nu_{e,h} (\sigma_+ k_-^{X} + \sigma_- k_+^{X} ).
\label{DresHam}
\end{equation}
Here $\beta _{e,h}$ are the effective Dresselhaus interaction constants of electrons and holes.

In addition to the SOIs, $k$-independent energy splitting between linearly polarized states, e.g. parallel and perpendicular to crystallographic axis can occur, and can be described as $k$-independent constant effective magnetic fields. Finally, we can write the Hamiltonian for indirect excitons on the basis of the four spin states $\left( {\begin{array}{*{20}{c}}
{{+2 }}&{{+1 }}&{{-1 }}&{{-2 }}
\end{array}} \right)^T$
as follows:
\begin{equation}
H = \left( {\begin{array}{*{20}{c}}
{{E_{ + 2}}\left( k^{X} \right)}&{\nu_e{\beta _e}{k_ +^{X} }}&{\nu_h{\beta _h}{k_ -^{X} }}&{ - {\delta _d}}\\
{\nu_e{\beta _e}{k_ -^{X} }}&{{E_{ + 1}}\left( k^{X} \right)}&{ - {\delta _b}}&{\nu_h{\beta _h}{k_ -^{X} }}\\
{\nu_h{\beta _h}{k_ +^{X} }}&{ - {\delta _b}}&{{E_{ - 1}}\left( k^{X} \right)}&{\nu_e{\beta _e}{k_ +^{X} }}\\
{ - {\delta _d}}&{\nu_h{\beta _h}{k_ +^{X} }}&{\nu_e{\beta _e}{k_ -^{X} }}&{{E_{ - 2}}\left( k^{X} \right)}
\end{array}} \right)
\label{Hamilton}
\end{equation}

Here $E_{\pm1,\pm2}(k)=E_{0}(k)=\hbar^2 (k^{X})^2/2m_{X}$ are the parabolic dispersions of the bare indirect exciton states, $\delta_{b,d}$ give the energy splittings between linearly polarized bright (dark) states. We take the exciton energy at $k^X=0$ as the zero point. Putting $\delta_b=\delta _d=0$, the Hamiltonian diagonalization is straightforward and yields the following set of isotropic eigenmodes:
\begin{equation}
\label{EigenValues}
\left( \begin{array}{c}
{E_I}\left( k^{X} \right)\\
{E_{II}}\left( k^{X} \right)\\
{E_{III}}\left( k^{X} \right)\\
{E_{IV}}\left( k^{X} \right)
\end{array} \right) = \left( {\begin{array}{*{20}{c}}
{{E_0}\left( k^{X} \right) + \left( {\nu_h{\beta _h} + \nu_e{\beta _e}} \right)k^{X}}\\
{{E_0}\left( k^{X} \right) + \left( {\nu_h{\beta _h} - \nu_e{\beta _e}} \right)k^{X}}\\
{{E_0}\left( k^{X} \right) - \left( {\nu_h{\beta _h} - \nu_e{\beta _e}} \right)k^{X}}\\
{{E_0}\left( k^{X} \right) - \left( {\nu_h{\beta _h} + \nu_e{\beta _e}} \right)k^{X}}
\end{array}} \right)\\
\end{equation}
and the corresponding eigenstates:
\begin{equation}\label{Eigenstates}
 \begin{array}{c}
{\psi _I} =\left( {\begin{array}{*{20}{c}}
1\\
{ + {e^{ - i\phi }}}\\
{ + {e^{ + i\phi }}}\\
1
\end{array}} \right),{\rm{ }}{\psi _{II}} = \left( {\begin{array}{*{20}{c}}
{ - 1}\\
{ + {e^{ - i\phi }}}\\
{ - {e^{ + i\phi }}}\\
1
\end{array}} \right)\\
{\psi _{III}} = \left( {\begin{array}{*{20}{c}}
{ - 1}\\
{ - {e^{ - i\phi }}}\\
{ + {e^{ + i\phi }}}\\
1
\end{array}} \right),{\rm{ }}{\psi _{IV}} = \left( {\begin{array}{*{20}{c}}
1\\
{ - {e^{ - i\phi }}}\\
{ - {e^{ + i\phi }}}\\
1
\end{array}} \right)
\end{array}
\end{equation}
Here $\phi$ is the polar angle in reciprocal space. The combinations of Eqs.\ref{Eigenstates} correspond to "linear polarizations" for the bright (rows 2 and 3) part given the usual identities $\psi_\pm=\psi_X\pm i\psi_Y$. While dark components are isotropic, the bright components have a linear polarization that is $\phi$-dependent and the polarization changes from $X$ to $Y$ when $\phi$ is changed by $\pi/2$. This peculiarity is analogue to the exciton-polariton case in the presence of the so-called TE-TM splitting that gives birth to the optical spin-Hall effect \cite{OSHEth,OSHEexp,OSHEPhoton,NOSHE}. One can therefore expect the formation of the same polarization pattern from the bright exciton states under proper excitation. 

\begin{figure}[ht]
  \includegraphics[width=0.4\textwidth]{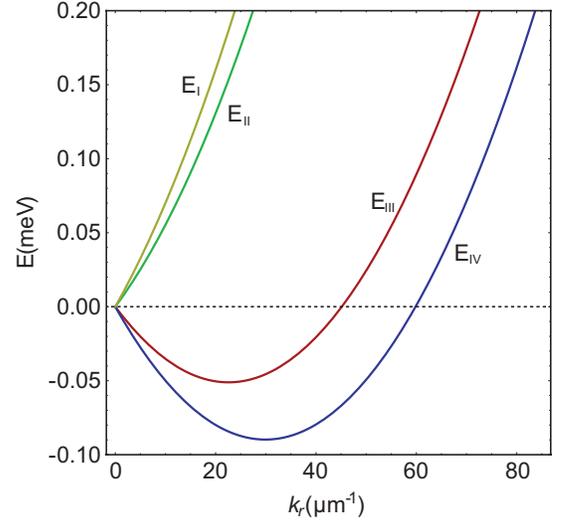}\\
  \caption{(Color online) Isotropic dispersion branches of the indirect exciton eigen modes [see Eqs.(\ref{EigenValues})].}
  \label{fig1}
\end{figure}

The isotropic dispersion branches given by the Eq.(\ref{EigenValues}) are plotted on the Fig.\ref{fig1} in the case $\delta_b=\delta_d=0$. One can clearly see the energy splittings between the branches coming from the Dresselhaus SOI contribution. Experimentally, one can detect only the emission coming from bright excitons with small wave-vectors  inside the light cone ($2.6\times10^7$m$^{-1}$). In this region the dispersion branches are linear. Interestingly, the calculated ground state of excitons has a significant wave vector completely out of the light cone. Thus a condensate in the ground state, even possessing some "bright" exciton component, is expected to be completely dark. It is moreover degenerate, and may demonstrate rich phenomenology of topological defects \cite{NowikBoltyk}.
The Liouville equation allows us to derive the reciprocal space dynamics of the spinor field $\boldsymbol{\Psi}=(\psi_{+2},\psi_{+1},\psi_{-1},\psi_{-2})^T$ with Eq.(\ref{Hamilton}) yielding:
\begin{equation}
i\hbar \frac{{\partial\boldsymbol{\Psi} }}{{\partial t}} = H\boldsymbol{\Psi}  - \frac{{i\hbar}}{2\tau }\boldsymbol{\Psi} + \mathbf{P}
\label{WaveEq}
\end{equation}
where we have phenomenologically introduced the exciton decay with a lifetime $\tau$ and a local source ${\bf{P}}({\bf{k^{X}}}) = {{\bf{A}}_P}\left( {\bf{k^{X}}} \right)\delta \left( {\omega  - {\omega _P}} \right)$  with frequency $\omega_P$ (in our calculations we took $\hbar \omega_P = 2 \mu eV$ ) acting on each component, where $\mathbf{A}_P(\mathbf{k^{X}})$ are 2D Gaussians. The formation of the condensate of indirect excitons coming from a hot reservoir and the process of their relaxation towards ground state are complex and need further investigations. Within this simplified model we consider a narrow pump spot (broad in reciprocal space) exciting laterally the dispersion branches and assume a ballistic propagation of the exciton. As said above, the ballistic propagation time is of order of a few ns, during which the evolution of the created cloud is expected to be coherent.

In the idealized case of a Dirac delta source in real space, the excitation of an eigenstate can be found analytically. For example, the dynamical equation for $\psi_I(k^{X})$ reads:
\begin{eqnarray}\label{psiI}
i\hbar \frac{{\partial {\psi _I}\left( {\mathbf{k^{X}},t} \right)}}{{\partial t}} = \frac{{{\hbar ^2}}}{{2m}}{(k^{X})^2}{\psi _I}\left( {\mathbf{k^{X}},t} \right) \nonumber\\
 + i\Delta \beta k^{X}{\psi _I}\left( {\mathbf{k^{X}},t} \right)+ {A_P}{e^{i{\omega _P}t}}\nonumber
\end{eqnarray}
and the stationary radial solutions are found writing $\psi_I(k_r^X,t)=\psi_I(k_r^X)\exp(i\omega_P t)$ which yields:
\begin{equation}\label{Stationary}
\hbar \omega_P {\psi _I}\left( {{k^{X}_r}} \right) = \left(\frac{{{\hbar ^2}}}{{2m}}{(k^{X}_r)^2} + i\Delta \beta k^{X}\right){\psi _I}\left( {{k^{X}_r}} \right) + {A_P}
\end{equation}
the Green's function of the problem reads
\begin{equation}\label{Green}
G_I\left( k^{X} \right) =  - \frac{{{A_P}}}{{\frac{{{\hbar ^2}}}{{2m}}(k^{X}_r)^2 + i\Delta \beta {k^{X}_r} - \hbar {\omega _P}}}
\end{equation}
whose Fourier transform gives
\begin{eqnarray}
\label{GreenReal1}
G_I\left( r \right) = \frac{{{A_P\kappa _1}}}{{{\kappa _2} - {\kappa _1}}}\left[ \begin{array}{l}
\log \left( {\frac{2}{{{\kappa _1}}}} \right)\frac{{{J_0}\left( {{\kappa _1}r} \right)}}{2} - \frac{\pi }{2}{\mathbf{H}_0}\left( {{\kappa _1}r} \right)\\
{ + _0}{F_1}\left( {1, - \frac{{\kappa _1^2{r^2}}}{4}} \right)
\end{array} \right]\\
\label{GreenReal2}
 - \frac{{{A_P\kappa _2}}}{{{\kappa _2} - {\kappa _1}}}\left[ \begin{array}{l}
\log \left( {\frac{2}{{{\kappa _2}}}} \right)\frac{{{J_0}\left( {{\kappa _2}r} \right)}}{2} - \frac{\pi }{2}{\mathbf{H}_0}\left( {{\kappa _2}r} \right)\\
{ + _0}{F_1}\left( {1, - \frac{{\kappa _2^2{r^2}}}{4}} \right)
\end{array} \right]
\end{eqnarray}
which is nothing but $\psi_I(r)$ for the delta source. $\kappa_{1,2}$ are the poles of Eq.(\ref{Green}) and Eqs.(\ref{GreenReal1},\ref{GreenReal2}) require that $\kappa_1>0$ and $\kappa_2<0$ which means that $\hbar \omega_P>0$. The solutions for the other eigenstates are found with a similar procedure and one can therefore construct any combination using the eigenstates (\ref{Eigenstates}).

\begin{figure}[ht]
  \includegraphics[width=0.4\textwidth]{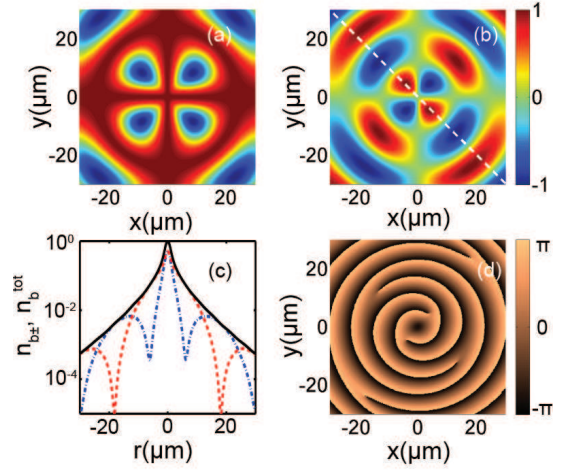}\\
  \caption{(Color online) (a) Degree of linear polarization of bright states. (b) Degree of circular polarization of bright states. (c) Density profile for $\sigma^{+}$- ($n_{b+}$, red dashed line),$\sigma^{-}$- ($n_{b-}$, blue dashed line) polarized bright states and total density ($n_{b}^{tot}$, black solid line) of bright states along the diagonal line $y=-x$ (white dashed line in (b)). (d) Phase of $\sigma^{+}$-polarized bright component. }
  \label{fig2}
\end{figure}

For a numerical simulation, we first consider a linearly polarised ($P_{\pm2}=0$, $P_{\pm1} \neq 0$) bright exciton state as initial condition. Here and after we took equal Dresselhaus constants for electrons and holes $\beta_{e,h}=6 \mu eV \cdot \mu m$, and the masses $m_e=0.07m_0$, $m_h=0.5m_0$, where $m_0$ is the free electron mass.The figure \ref{fig2}(a,b,c) shows the stationary real space images of a linear, circular polarization degree and phase of bright states.  One can observe the formation of a Skyrmion lattice associated with the formation of the spin domains. The situation is analogous with the
polaritonic spin Hall effect recently analyzed theoretically \cite{NOSHE}. This similarity is expectable, since the effective field acting on the bright states is exactly equivalent to the case of exciton polaritons in the presence of the so-called TE-TM splitting. Fig.\ref{fig2} (c) shows the density profile of bright states along the diagonal line [white dashed line on Fig.\ref{fig2}(b)]. Densities of $\sigma^{+}$ and $\sigma^{-}$ states are oscillating while the total density decays in space with $r$. Phase structure of $\sigma^{+}$-polarized bright state is plotted on Fig.\ref{fig2}(d).
However, the linear polarization pattern that we observe (8 polarization domains), differs from the one measured in \cite{ButovNat2012} (4 domains).

In order to reproduce the experiment, we therefore consider a different initial spin state for the condensate (Fig. 3). We consider first a condensate of dark states with a slight asymmetry between dark components (${P_{-2}}/{P_{+2}}=0.9$, $P_{+1}=P_{-1} = 0$). Then, in order to mix the circularly polarized bright states, we introduce an additional constant splitting between the linearly polarized bright states along and perpendicular to the main crystallographic axis of the sample $\delta_{b}$ = $1 \mu eV$.

\begin{figure}[ht]
  \includegraphics[width=0.4\textwidth]{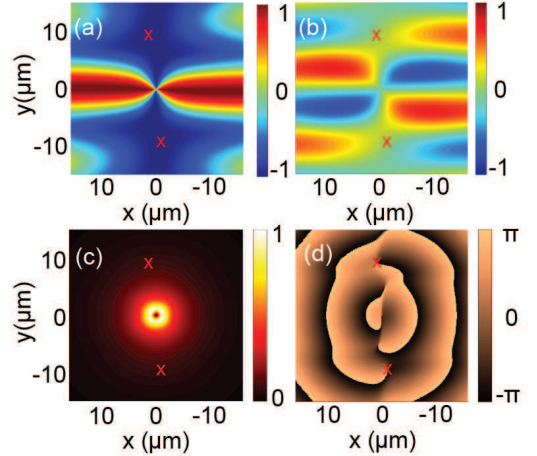}\\
  \caption{(Color online) (a) Degree of linear polarization of bright states. (b)Degree of circular polarization of bright states. (c) Total density of bright states. (d) Phase of $x$-component of bright states. Red crosses point the phase singularities in $x$-component}
  \label{fig3}
\end{figure}

Fig. \ref{fig3} (a,b) shows the spatial distribution of linear and circular polarization degree, while Fig.\ref{fig3}(d) shows the phase of the $X$-component. Both circular and linear polarization structures present 4 domains and are stretched along the $x$ axis (because of $\delta_b$) as it was observed in experiment\cite{ButovNat2012}. Interestingly, we also observe pairs of phase singularities (red crosses), situated symmetrically with respect to the exciton source. This phase singularity is accompanied by a density dip only in the $Y$ component. This topological defect is therefore similar to a Skyrmion but in the linear polarization basis. Our approach shows that the appearance of phase singularities is a general feature of radial flows of particles in the presence of coupling between the spin and motional degree of freedom.

\emph{Conclusions}.
We have shown that an expanding  cloud of indirect excitons can form polarization textures and phase singularities thanks to the interplay between the flow and spin-orbit interactions. Their structure strongly depends on the polarization of initial state, so changing the excitation conditions one can strongly modify the number of polarization domains and qualitatively reproduce experimental data. Additionally, we were able to mimic the optical spin-Hall effect for bright excitons.

\emph{Acknowlegements}.
This work was supported by EU ITNs "Spin-Optronics" Grant No. 237252 and INDEX Grant No. 289968. Also we would like to thank L. Butov and H. Tercas for useful discussions.

\bibliographystyle{apsrev4-1}
\bibliography{ref}

\end{document}